\documentclass[aps,prl,twocolumn,superscriptaddress,showpacs]{revtex4-1}
\usepackage[colorlinks,citecolor=blue,urlcolor=blue,linkcolor=blue,bookmarks=false,hypertexnames=true]{hyperref}
\usepackage{xcolor}
\usepackage{multirow}
\usepackage[normalem]{ulem}

\usepackage{graphicx}  
\graphicspath{{figures/}} 
\usepackage{dcolumn}   
\usepackage{bm}        
\usepackage{amssymb}   
\usepackage{ulem}
\usepackage{soul}
\usepackage{makecell}

\colorlet{mycolor}{black}
\colorlet{mycolor2}{black}

\begin{document}
\title{Multiple Mechanisms in Proton-Induced Nucleon Removal at $\sim$100 MeV/Nucleon}

\newcommand{\akth}{        \affiliation{Department of Physics, Royal Institute of Technology, SE-10691 Stockholm, Sweden}}
\newcommand{\aatomki}{     \affiliation{MTA Atomki, P.O. Box 51, Debrecen H-4001, Hungary}}
\newcommand{\abeijing}{    \affiliation{State Key Laboratory of Nuclear Physics and Technology, Peking University, Beijing 100871, P.R. China}}
\newcommand{\acaen}{       \affiliation{LPC Caen, ENSICAEN, Université de Caen, CNRS/IN2P3, F-14050 Caen, France}}
\newcommand{\acea}{        \affiliation{IRFU, CEA, Universit\'e Paris-Saclay, F-91191 Gif-sur-Yvette, France}}
\newcommand{\acns}{        \affiliation{Center for Nuclear Study, University of Tokyo, RIKEN campus, Wako, Saitama 351-0198, Japan}}
\newcommand{\aewha}{       \affiliation{Department of Physics, Ewha Womans University, Seoul, South Korea}}
\newcommand{\agsi}{        \affiliation{GSI Helmoltzzentrum f\"ur Schwerionenforschung GmbH, Planckstr. 1, 64291 Darmstadt, Germany}}
\newcommand{\ahku}{        \affiliation{Department of Physics, The University of Hong Kong, Pokfulam, Hong Kong}}
\newcommand{\ainst}{       \affiliation{Institute for Nuclear Science \& Technology, VINATOM, 179 Hoang Quoc Viet, Cau Giay, Hanoi, Vietnam}}
\newcommand{\aipno}{       \affiliation{Université Paris-Saclay, CNRS/IN2P3, IJCLab, 91405 Orsay, France}}
\newcommand{\akoeln}{      \affiliation{Institut f\"ur Kernphysik, Universit\"at zu K\"oln, D-50937 Cologne, Germany}}
\newcommand{\alanzhou}{    \affiliation{Institute of Modern Physics, Chinese Academy of Sciences, Lanzhou, China}}
\newcommand{\amadrid}{     \affiliation{Instituto de Estructura de la Materia, CSIC, E-28006 Madrid, Spain}}
\newcommand{\aorsay}{      \affiliation{CSNSM, CNRS/IN2P3, Universit\'e Paris-Sud, F-91405 Orsay Campus, France}}
\newcommand{\aoslo}{       \affiliation{Department of Physics, University of Oslo, N-0316 Oslo, Norway}}
\newcommand{\ariken}{      \affiliation{RIKEN Nishina Center, 2-1 Hirosawa, Wako, Saitama 351-0198, Japan}}
\newcommand{\arikkyo}{     \affiliation{Department of Physics, Rikkyo University, 3-34-1 Nishi-Ikebukuro, Toshima, Tokyo 172-8501, Japan}}
\newcommand{\atitech}{     \affiliation{Department of Physics, Tokyo Institute of Technology, 2-12-1 O-Okayama, Meguro, Tokyo, 152-8551, Japan}}
\newcommand{\atohoku}{     \affiliation{Department of Physics, Tohoku University, Sendai 980-8578, Japan}}
\newcommand{\atohokuc}{     \affiliation{Cyclotron and Radioisotope Center, Tohoku University, Sendai 980-8578, Japan}}
\newcommand{\akonan}{     \affiliation{Department of Physics, Konan University, Kobe 658-8501, Japan}}
\newcommand{\atud}{        \affiliation{Institut f\"ur Kernphysik, Technische Universit\"at Darmstadt, 64289 Darmstadt, Germany}}
\newcommand{\aunal}{       \affiliation{Universidad Nacional de Colombia, Carr. 30 No. 45-03, Bogot\'a, Colombia}}
\newcommand{\atokyo}{      \affiliation{Department of Physics, University of Tokyo, 7-3-1 Hongo, Bunkyo, Tokyo 113-0033, Japan}}
\newcommand{\aoak}{        \affiliation{Physics Division, Oak Ridge National Laboratory, Oak Ridge, Tennessee 37831, USA}}
\newcommand{\atennessee}{  \affiliation{Department of Physics and Astronomy, University of Tennessee, Knoxville, Tennessee 37996, USA}}
\newcommand{\atriumf}{     \affiliation{TRIUMF 4004 Wesbrook Mall, Vancouver, British Columbia V6T 2A3, Canada}}
\newcommand{\amaxplank}{   \affiliation{Max-Planck-Institut f\"ur Kernphysik, Saupfercheckweg 1, 69117 Heidelberg, Germany}}
\newcommand{\arcnp}{       \affiliation{Research Center for Nuclear Physics (RCNP), Osaka University, Ibaraki 567-0047, Japan}}
\newcommand{\ajaea}{       \affiliation{Advanced Science Research Center, Japan Atomic Energy Agency, Tokai, Ibaraki 319-1195, Japan}}
\newcommand{\asurrey}{     \affiliation{Department of Physics, University of Surrey, Guildford GU2 7XH, UK}}
\newcommand{\aku}{         \affiliation{KU Leuven, Instituut voor Kern- en Stralingsfysica, 3001 Leuven, Belgium}}
\newcommand{\astrasbourg}{ \affiliation{Universit\' de Strasbourg, IPHC, 67037 Strasbourg Cedex, France}}
\newcommand{\atokuyama}{   \affiliation{National Institute of Technology, Tokuyama College, Shunan 745-8585, Japan}}
\newcommand{\apku}{        \affiliation{State Key Laboratory of Nuclear Physics and Technology, School of Physics, Peking University, Beijing 100871, China}}
\newcommand{\amiyazaki}{   \affiliation{Department of Applied Physics, University of Miyazaki, Gakuen-Kibanadai-Nishi 1-1, Miyazaki 889-2192, Japan}}
\newcommand{\akyushu}{     \affiliation{Department of Physics, Kyushu University, Fukuoka 812-8581, Japan}}
\newcommand{\atum}{        \affiliation{Department of Physics, Technische Universit\"at M\"unchen, James-Franck-Stra{\ss}e 1, 85748 Garching, Germany}}
\newcommand{\akyoto}{      \affiliation{Department of Physics, Kyoto University, Kitashirakawa, Sakyo, Kyoto 606-8502, Japan}}
\newcommand{\ausa}{      \affiliation{Departamento de F\'isica At\'omica, Molecular y Nuclear, Facultad de F\'isica, Universidad de Sevilla, Apartado 1065, E-41080 Sevilla, Spain}}
\newcommand{\azsd}{        \affiliation{Sino-French Institute of Nuclear Engineering and Technology, Sun Yat-Sen University, Zhuhai, 519082, Guangdong, China}}
\newcommand{\abnu}{        \affiliation{Key Laboratory of Beam Technology and Material Modification of Ministry of Education,
College of Nuclear Science and Technology, Beijing Normal University, Beijing 100875, China}}
\newcommand{\anscl}{        \affiliation{Department of Physics and Astronomy and the Facility for Rare Isotope Beams, Michigan State University, East Lansing, Michigan 48824-1321, USA }}
\newcommand{\acens}{        \affiliation{Center for Exotic Nuclear Studies, Institute for Basic Science, Daejeon 34126, Republic of Korea}}
\newcommand{\ahoria}{       \affiliation{Horia Hulubei National Institute for R$\&$D in Physics and Nuclear Engineering, IFIN-HH, 077125 Bucure\c{s}ti-M\u{a}gurele, Romania}}
\newcommand{\abuch}{        \affiliation{Doctoral School of Physics, University of Bucharest, 077125 Bucure\c{s}ti-M\u{a}gurele, Romania}}

\author{T.~Pohl} \atud
\author{Y.~L.~Sun} \email{ysun@ikp.tu-darmstadt.de} \atud \acea \ahku
\author{A.~Obertelli} \atud \acea
\author{J.~Lee} \ahku
\author{M. G\'omez-Ramos} \ausa
\author{K.~Ogata} \akyushu \arcnp
\author{K.~Yoshida} \ajaea
\author{B.~S.~Cai}  \azsd
\author{C.~X.~Yuan}  \azsd
\author{B.~A. Brown}  \anscl
\author{H.~Baba} \ariken
\author{D.~Beaumel} \aipno
\author{A.~Corsi}\acea
\author{J.~Gao} \ariken \apku
\author{J.~Gibelin} \acaen
\author{A.~Gillibert} \acea
\author{K.~I.~Hahn} \aewha \acens
\author{T.~Isobe} \ariken
\author{D.~Kim} \aewha \acens
\author{Y.~Kondo}\atitech
\author{T.~Kobayashi} \atohoku
\author{Y.~Kubota} \ariken \acns
\author{P.~Li} \ahku
\author{P.~Liang} \ahku
\author{H.~N.~Liu} \atud \acea \abnu 
\author{J.~Liu} \ahku
\author{T.~Lokotko} \ahku
\author{F.~M.~Marqu\'{e}s} \acaen
\author{Y.~Matsuda} \atohokuc \akonan
\author{T.~Motobayashi} \ariken
\author{T.~Nakamura} \atitech
\author{N.~A.~Orr} \acaen
\author{H.~Otsu} \ariken
\author{V.~Panin} \ariken \acea
\author{S.~Y.~Park} \aewha \ariken
\author{S.~Sakaguchi} \akyushu
\author{M.~Sasano} \ariken
\author{H.~Sato} \ariken
\author{H.~Sakurai} \ariken \atokyo
\author{Y.~Shimizu} \ariken
\author{A.~I.~Stefanescu} \ahoria \abuch \ariken
\author{L.~Stuhl} \ariken \acens
\author{D.~Suzuki} \ariken
\author{Y.~Togano} \atitech \ariken \arikkyo
\author{D.~Tudor} \ahoria \abuch \ariken
\author{T.~Uesaka} \ariken
\author{H.~Wang} \ariken
\author{X.~Xu} \ahku
\author{Z.~H.~Yang} \ariken
\author{K.~Yoneda} \ariken
\author{J.~Zenihiro} \ariken

\date{\today}

\begin{abstract}
   We report on the first proton-induced single proton- and neutron-removal reactions from the neutron-deficient $^{14}$O nucleus with large 
   Fermi-surface
   asymmetry $S_n-S_p$ = 18.6~MeV at $\sim$100 MeV/nucleon,
   a widely used energy regime for rare-isotope studies.
   The measured inclusive cross sections and parallel momentum distributions of the $^{13}$N and $^{13}$O residues are compared to the state-of-the-art reaction models, with nuclear structure inputs from many-body shell-model calculations.
   Our results provide the first quantitative contributions of multiple reaction mechanisms including the quasifree knockout, inelastic scattering and nucleon transfer processes.
   It is shown that the inelastic scattering and nucleon transfer, usually neglected at such energy regime, contribute about 50\% and 30\% to the loosely bound proton and deeply bound neutron removal, respectively. 
   These multiple reaction mechanisms should be considered in analyses of inclusive one-nucleon removal cross sections measured at intermediate energies for quantitative investigation of single-particle strengths and correlations in atomic nuclei.
\end{abstract}

\maketitle

The concept of independent particle motion has played a fundamental role for the study of quantum many-body systems, such as metallic clusters, atoms and nuclei \cite{Pandharipande1997}.
For the nuclear many-body systems, a first-order description was realized via the independent particle model \cite{mayer49,haxeljensensuess49,Bohr1969}, 
in which nucleons move freely in an effective mean-field potential provided by all the other nucleons.
Nucleon-nucleon correlations should be added for a realistic description of nuclear properties.
It was first revealed by ($\textit{e}$,$\textit{e}'\textit{p}$) experiments on stable nuclei that the nuclear single-particle strengths, 
quantified by the so-called spectroscopic factors, are reduced by (30--40)\% relative to the independent particle model predictions \cite{lapikas93, Kramer2001}. 
The ``quenching" of the single-particle strengths has been attributed to long-range \cite{dickhoff04,Barbieri2009} and short-range \cite{Piasetzky2006,Subedi2008,Hen2014, Duer2018} correlations, 
whose investigations have significantly improved our understanding of the \textcolor{mycolor}{strongly} correlated nuclear many-body system and led to new insights for dense nuclear matter such as neutron stars \cite{Ding2016,Fomin2017}.
One main focus of today’s nuclear physics is to extend the correlation studies toward the proton and neutron driplines \cite{Paschalis2020,Aumann2021}.

One-nucleon removal reactions at intermediate energies near and above 100 MeV/nucleon 
have been a powerful tool to extract single-particle strengths of unstable nuclei~\cite{Hansen2003}.
The quenching of the single-particle strengths has been connected to the so-called reduction factor $R_{s}$~\cite{Brown2002}, defined as the ratio of the experimental to the theoretical cross section that is usually computed using shell-model spectroscopic factors and a reaction model \textcolor{mycolor}{relying on the adiabatic (or sudden) and eikonal approximations~\cite{Hansen2003}.}
\textcolor{mycolor}{
The central assumptions are that the residue and removed-nucleon relative motion is considered as frozen and the trajectories follow straight lines before and after the collision.}
Systematic studies from light-ion-induced one-nucleon removal reactions at $\sim$100 MeV/nucleon \cite{Gade2008,tostevingade14,tostevingade21} and higher incident beam energies~\cite{Sun2022} revealed that {\it R}$_{s}$ has a strong dependence on the Fermi-surface asymmetry quantified as $\Delta S$ = $S_n$ - $S_p$ or $S_p$ - $S_n$ for neutron or proton removal, respectively.
However, results from transfer reactions~\cite{lee06,lee10,flavigny13, Kay2013,Xu2019,Manfredi2021,Kay2022} and proton-induced quasifree knockout $(p, pN)$ reactions~\cite{Kawase2018,Atar2018, Gomez2018, Holl2019, Nguyen2019} did not confirm the strong $\Delta S$ dependence.

The inconsistent dependence on $\Delta S$ calls for a deeper understanding of reaction mechanisms and correlations in nuclei~\cite{Aumann2021}.
Although the diffraction and stripping mechanisms have been well-established in the eikonal model down to $\sim$100 MeV/nucleon~\cite{Bazin2009,Wimmer2014},
multiple scattering, excitation and decay of the one-nucleon removal residue, beyond the eikonal reaction model ~\cite{Louchart2011, Sun2016} or Pauli blocking ~\cite{Flavigny2009,Bertulani2010} have been proposed as possible mechanisms that could reduce the deeply bound nucleon-removal cross sections. 
In particular, asymmetric parallel momentum distributions (PMDs) of the residue, characterized by a low-momentum tail ~\cite{Tostevin2002,Gade2005,Yurkewicz2006,Grinyer2011,flavigny12,Shane2012,Stroberg2014,Charity2020} and a high-momentum cutoff~\cite{flavigny12},
have been observed in several experiments,
in contrast to the symmetric PMDs predicted by the lowest order eikonal model~\cite{Bertulani1992,Tostevin2001,Aumann2013}. 

\begin{figure}[!h]
    \centering
     \includegraphics[width=8.0cm, height=7.5cm]{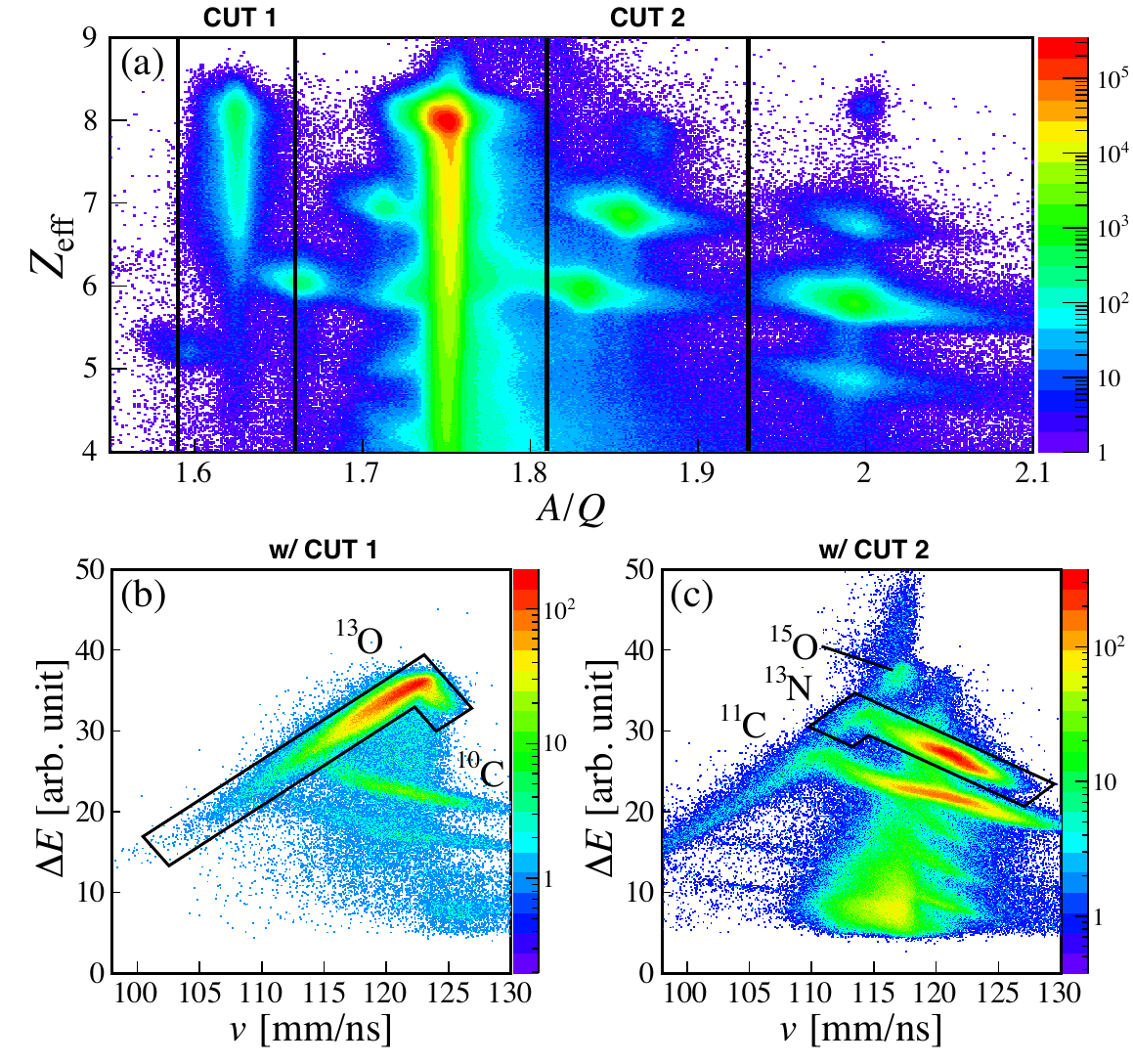}
    \caption{Particle identification of projectilelike residues.
    \textcolor{mycolor}{Particle's velocity ($\nu$) is deduced from TOF. Combining $B\rho$ and velocity allows to determine particle's mass-to-charge ratio $\textit{A}$/$\textit{Q}$. $\textit{Z}\rm_{eff}$ is the deduced effective atomic number with $\Delta E$ and velocity using the Bethe-Bloch formula. The energy deposit of a charged particle in a material is related to its velocity, charge and mass, as shown in the $\Delta E$-$\nu$ spectra with (w/) the $A/Q$ selections} from 1.59 to 1.66 (b) and from 1.81 to 1.93 (c). Black contours in (b) and (c) select $^{13}$O and $^{13}$N, respectively.}
    \label{fig:PID_selection} 
\end{figure}

Aforementioned studies~\cite{Gade2008,tostevingade14,tostevingade21,Sun2022} have been conducted with
light absorptive nuclear targets, $^9$Be or $^{12}$C,
which introduce the complexity that the final state of the target is unknown. 
Here, we report on the first study of one-nucleon removal from a large Fermi-surface asymmetric nucleus $^{14}$O ($\Delta S$ = $\pm$18.6 MeV) at $\sim$100~MeV~/~nucleon using a single-nucleon target, i.e., protons. 
$^{14}$O is an ideal nucleus to study the one-nucleon removal mechanisms at large proton-to-neutron asymmetry.
The proton and neutron removal from it involves only the orbitals of $\pi$0$p_{1/2}$ and $\nu$0$p_{3/2}$, respectively, since both $^{13}$N ($J^{\pi}\rm_{g.s.}$ = 1/2$^{-}$) and $^{13}$O ($J^{\pi}\rm_{g.s.}$ = 3/2$^{-}$) do not exhibit bound excited states.  
Based on the measured PMDs and the state-of-the-art reaction models, we show that in addition to the quasifree knockout, the inelastic scattering and nucleon transfer also make significant contributions to the loosely bound proton removal and deeply bound neutron removal, respectively.

The experiment was performed at the Radioactive Isotope Beam Factory operated by the RIKEN Nishina Center and the Center for Nuclear Study, The University of Tokyo. A primary $^{18}$O beam at 230\,MeV/nucleon with an intensity of 500\,pnA bombarded on a 14-mm thick $^{9}$Be target.
The $^{14}$O secondary beam was purified and identified using the time of flight (TOF) and the energy loss ($\Delta E$) information by the BigRIPS fragment separator~\cite{Fukuda13}.
\begin{table}[thpb]
    \centering
    \caption{Experimental ($\sigma\rm_{exp}$) and theoretical ($\sigma\rm_{th}$) cross sections for one-nucleon removal from $^{14}$O at 94\,MeV/nucleon. 
    SF represents the spectroscopic factor from shell-model calculations (see SM~\cite{SM}).
The reduction factors $R_s$ = $\sigma_{\text{exp}}$/$\sigma_{\text{th}}$ are also given.}
    \begin{tabular}{|c|c|c|c|ccc|c|}
            \hline
         Residue & J$^{\pi}$ &$\sigma_{\text{exp}}$& SF & Theory &  $\sigma_{\text{sp}}$& $\sigma_{\text{th}}$& $R_{\text{s}}$ \\ 
        & & [mb] & & & [mb] & [mb] & \\
         \hline
        $^{13}$N$ \rm_{g.s.} $  & 1/2$^{-}$& 10.7(16)      & 1.58      & DWIA   & 5.2  & 8.8  & 1.22(18) \\
                                        & &          &  & Inelastic   &    -  & 9    &  \\  
                                        & &          &        &Sum    &      & 17.8 & 0.60(9) \\
                                            \cline{5-8}
                                        & &          &         & QTC   & 7.0 & 11.9 & 0.90(13) \\ 
                                       &  &          &  & Inelastic   &   -   & 9    &  \\ 
                                       &  &          &        & Sum   &      & 20.9 & 0.51(8) \\ \hline 
         $^{13}$O$\rm_{g.s.}$   &  3/2$^{-}$& 16.7(24)    & 3.42       & DWIA  & 6.3 & 23.2 &0.72(10) \\ 
                                           & &          &    & Transfer  & 3 & 11 & \\ 
                                          &  &          &         & Sum  & & 34.2 & 0.49(7) \\
                                            \cline{5-8}
                                        &    &          &          & \makecell{QTC \\ w/o transfer}  & 10.2 & 37.6  & 0.44(6)\\
                                        &    &          &          & QTC  & 13.5 & 49.7  & 0.34(5)\\\hline
    \end{tabular}
    \label{tab:crosssections}
\end{table}
The typical $^{14}$O beam intensity and purity were 9 $\times$ 10$^{3}$ particles per second and 78\%, respectively.
The $^{14}$O beam was tracked onto a \textcolor{mycolor}{2.40(34)-mm thick} solid hydrogen target \cite{Matsuda2011} using two multiwire drift chambers.
The beam energy at the target center was 94\,MeV/nucleon with a narrow spread of 0.2 MeV/nucleon ($\sigma$).
The target density was determined to be 86 mg/cm$^{3}$ based on the monitored target-cell temperature.
The target thickness and its uncertainty were extracted by measuring the momentum change of the unreacted $^{14}$O beam with and without the hydrogen target.
The empty-target setting was also used to measure the background generated by nontarget beamline materials, which were subtracted in the cross section and PMD analyses.

The reaction residues were measured by the SAMURAI spectrometer~\cite{Kobayashi2013}, 
with a magnetic field set at 1.49\,Tesla with filled target and 1.51\,Tesla with empty target. 
Positions and angles of the particles were measured by two multiwire drift chambers located before and after the dipole magnet.
A 10-mm thick plastic scintillator array hodoscope located downstream of the spectrometer was used to measure the $\Delta E$ and to determine the TOF together with a 0.2-mm thick plastic scintillator before the target.
The magnetic rigidity $B\rho$ and the flight length from the target to hodoscope were deduced from multidimensional-fit functions
using measured positions and angles as inputs.
The functions, obtained through Geant4 \cite{Agost2003} simulations and multidimensional fit package of ROOT \cite{Brun1997},
reproduce the simulated $B\rho$ and flight length with relative deviations below 0.02\,\%.

As shown in Fig.~\ref{fig:PID_selection},
$^{13}$O and $^{13}$N can be unambiguously identified using the $\Delta E$-$B\rho$-TOF and $\Delta E$-velocity method.
In Fig.~\ref{fig:PID_selection}(a), the deduced atomic number $\textit{Z}\rm_{eff}$ for $^{13}$O ($A$/$Q$ = 1.625) and $^{14}$O ($A$/$Q$ = 1.75) both show tails extending to smaller $\textit{Z}\rm_{eff}$ region. The $\textit{Z}\rm_{eff}$ tail of $^{14}$O is caused by unreacted $^{14}$O projectiles interacting in the hodoscope, while the $\textit{Z}\rm_{eff}$ tail of $^{13}$O has a strong component steaming from the low-energy $^{13}$O stopped in the hodoscope. 
As demonstrated in Fig.~\ref{fig:PID_selection}(b) and (c), most $^{13}$O stopped in the hodoscope and had $\Delta E$ proportional to velocity, while most $^{13}$N punched through the hodoscope and had $\Delta E$ antiproportional to velocity.

\begin{figure}[h]
    \centering
    \includegraphics[trim=0cm 0.2cm 0.8cm 1.2cm,clip,width=8.5cm,height=8.0cm]{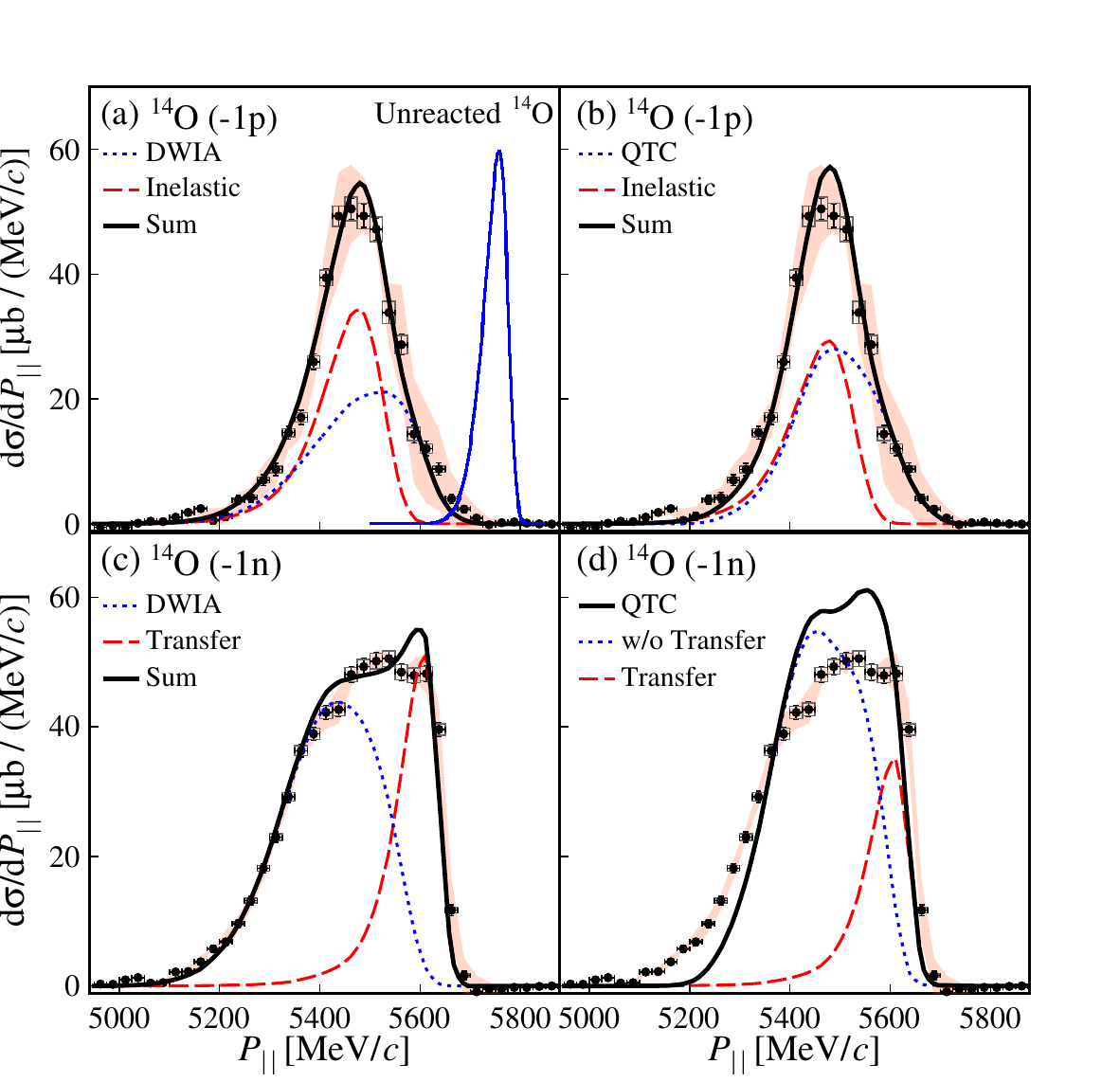}
    \caption{Parallel momentum distributions of $^{13}$N and $^{13}$O.
    The black-filled markers show the experimental data. The orange bands represent the uncertainties from the background subtractions. The gray empty bins indicate the other systematic uncertainties. 
    Panels (a)-(c) compare the data to DWIA and QTC calculations (blue dotted lines), 
    with additional contributions from inelastic excitation for $^{13}$N and ($\textit{p}$, $\textit{d}$) transfer for $^{13}$O (red-dashed lines). 
    \textcolor{mycolor}{Panel (d) displays the QTC calculation as a black solid line, while the calculation without (w/o) ($\textit{p}$, $\textit{d}$) transfer is represented by a blue dotted line.}
    The blue solid line in (a) shows the distribution of the unreacted $^{14}$O beam (shifted by $-$200\,MeV/$c$) to demonstrate the experimental response,
    \textcolor{mycolor}{which introduces a shift and broadening of the momentum due to energy losses and detector resolutions.}
    Theoretical distributions have been convoluted with the experimental response and their
    integrals have been normalized \textcolor{mycolor}{to the experimental cross sections, without any momentum shift to match the data. \textcolor{mycolor2}{Note that the sharp ($\textit{p}$, $\textit{d}$) transfer peak is smoothed out by the experimental response. See SM~\cite{SM} for theoretical distributions before experimental response convolution.}}
    }
    \label{fig:PMD} 
\end{figure}

The resulting experimental cross sections 
are listed in Table.~\ref{tab:crosssections}. 
Momentum acceptances, 94(1)\% for $^{13}$O and 96(1)\% for $^{13}$N, determined from Geant4 simulations and 7(1)\% reaction loss in the beamline materials have been taken into account.
In addition, a 5(1)\,\% loss was considered to account for
$^{13}$O or $^{13}$N events outside of the gates in Figs.~\ref{fig:PID_selection} (b) and (c),
based on simulations with the
\textcolor{mycolor}{Li\`{e}ge Intranuclear Cascade model~\cite{incl}, which is a well-established model for the description of spallation reactions.}
The cross section errors for $^{13}$O and $^{13}$N contain statistical uncertainties (0.6\,\% and 1.3\,\%), particle selections (0.9\,\% and 2.3\,\%) and systematic uncertainties (14.2\,\% and 14.7\,\%) mainly resulting from the target-thickness uncertainty.
The experimental PMDs of $^{13}$O and $^{13}$N are shown in Fig.~\ref{fig:PMD}. 
An asymmetric PMD with a low-momentum tail and a high-momentum sharp edge is observed in the deeply bound neutron removal channel, 
while the PMD from the loosely bound proton removal is close to symmetric.

The experimental cross sections and PMDs were compared to predictions combining structure and reaction inputs. 
Spectroscopic factors for the removed nucleons were obtained from shell-model calculations. 
See Supplemental Material (SM)~\cite{SM} for details.
For the ($\textit{p}$, $\textit{pN}$) knockout process, we adopted the DWIA (Distorted-Wave Impulse Approximation)~\cite{Chant1977, Chant1983,ogata15,Wakasa2017,Yoshida2018} and the QTC (Quantum Transfer-to-the-Continuum)~\cite{Moro2015,Gomez2018} models (see SM~\cite{SM}).
\textcolor{mycolor}{Both models assume a single scattering between the removed nucleon and the target proton, using the transition amplitude to calculate the cross sections. The key disparity lies in how they handle the three-body final state: DWIA factorizes it as the product of the $\textit{p}$-residue and $\textit{N}$-residue states, while QTC expands it in terms of $\textit{p}$-$\textit{N}$ states,
including deuteron ground state for neutron removal~\cite{Yoshida2018}.}
The DWIA and QTC reaction models have been developed and benchmarked for ($\textit{p}$, $\textit{pN}$) reactions at beam energies higher than 200 MeV/nucleon~\cite{Gomez2018,Yoshida2018}. 
Above 200 MeV/nucleon, both models reproduce well the shape of the experimental momentum distributions~\cite{Sun2020,Chen2019,Browne2021,Gomez2018}, 
and the calculated single-particle cross sections ($\sigma\rm_{sp}$) are consistent with each other within~20\%~\cite{aumann2020}.
The obtained $\sigma\rm_{sp}$ for $^{14}$O($\textit{p}$, 2$\textit{p}$)$^{13}$N and 
$^{14}$O($\textit{p}$, $\textit{pn}$)$^{13}$O reactions are listed in Table.~\ref{tab:crosssections}.

In addition, we also considered the ($\textit{p}$, $\textit{p}'$) inelastic excitation of $^{14}$O 
to its low-lying excited states located above $S_{p}$ and below $\sim S_{2p}$, which decay to the ground state of $^{13}$N via one-proton emission.
Giant-resonance excitations were not considered.
A total inelastic cross section of 9 mb was obtained (see SM~\cite{SM}).
As shown in Figs.~\ref{fig:PMD} (a) and (b), the sum of the ($\textit{p}$, 2$\textit{p}$) and ($\textit{p}$, $\textit{p}'$) PMDs is close to symmetric and reproduces well the PMD of $^{13}$N.
The good agreement confirms the predicted strong inelastic-scattering component in the loosely bound proton removal, which has fractional contributions of 51\% with the DWIA and of 43\% with the QTC.
The inelastic-scattering component was also observed in the one-nucleon removal with a $^{9}$Be target using the invariant-mass technique~\cite{Charity2019, Charity2020}.
Percentage contributions of 17\% and 21\% from the inelastic scattering were extracted for the one-proton removal from $^{9}$C and $^{13}$O at $\sim$65 MeV/nucleon~\cite{Charity2020}.
If the inelastic-scattering component is ignored, the present one-proton removal {\it R}$_{s}$ will be around unity, 
coinciding with the loosely bound nucleon-removal {\it R}$_{s}$ from eikonal model based analysis \cite{Bazin2009,Gade2008,tostevingade14,tostevingade21}. 
\textcolor{mycolor}{Performing additional coincidence measurements with the recoil and decayed protons to obtain the angular distribution and the $^{14}$O excitation energy would further characterize the inelastic components.}
The low-lying excited states considered here have multiparticle-multihole configurations.
It was shown recently that inelastic scattering with large momentum transfer has the advantage of populating multiparticle-multihole states~\cite{Gade2022}.
Descriptions of such states are beyond the $(p,pN)$ and the eikonal models, which assume beforehand that the projectile is a single-particle state plus an inert core~\cite{Hansen2003}.

For the deeply bound neutron removal, 
the $(p, d)$ transfer is considered in the QTC formalism but not in the DWIA.
To study the ($\textit{p}$, $\textit{d}$) transfer effect, we performed the QTC calculation with the outgoing channel coupled only to the deuteron ground state, that is equivalent to the so-called DWBA (Distorted-Wave Born Approximation) calculation.
The obtained $\sigma\rm_{sp}$ for the transfer reaction is 3 mb \textcolor{mycolor}{with an uncertainty of about 1 mb (see SM~\cite{SM}).}
The QTC $\sigma\rm_{sp}$ without the ($\textit{p}$, $\textit{d}$) transfer is still larger than the DWIA result. 
Other effects, such as low-energy neutron-core absorption, contribute to this difference.

As shown in Fig.~\ref{fig:PMD} (c),
the PMD of $^{13}$O is well reproduced by combining the contributions from the DWIA and the DWBA,
in which the latter corresponding to $(p, d)$ transfer contributes to $\sim$30\%. 
Our data supports the interpretation of Ref.~\cite{ogata15} that the low-momentum tail is caused by the attractive potential between the outgoing nucleons and $^{13}$O. 
Meanwhile, the ($\textit{p}$, $\textit{d}$) transfer reaction creates a sharp high-momentum edge, as observed in the data, due to the two-body kinematics of the transfer reaction. 
The sharp edge is found in a kinematic region inaccessible to ($\textit{p}$, $\textit{pn}$) knockout and is thus a proof of the significant transfer contribution.
\textcolor{mycolor}{Note that the sharp edge here has a different origin with that observed in Ref.~\cite{flavigny12}, which is due to a threshold effect when the incident energy per particle is comparable to the nucleon separation energy. 
Additional characterization of the transfer contribution would be to measure the angular distribution of cross sections in coincidence with the deuteron. See for example Ref.~\cite{Kasagi1983}.}
Since QTC formalism treats ($\textit{p}$, $\textit{d}$) transfer consistently with ($\textit{p}$, $\textit{pn}$), it reproduces better the sharp high-momentum side than DWIA, as shown in Fig.~\ref{fig:PMD} (d).
However, QTC does not reproduce the low-momentum tail as well as DWIA. The reason might be due to the different treatment of the final state interaction in QTC, especially that the nucleon-residue interaction at low relative energy is not explicitly treated in QTC formalism.

It is the first time the PMD measured near 100 MeV/nucleon shows a distinctive contribution from the ($\textit{p}$, $\textit{d}$) transfer reaction, usually neglected at such beam energies~\cite{Aumann2021}.
One-nucleon pickup cross sections have been measured around 60 MeV/nucleon with heavy-ion beams on $^{12}$C or $^9$Be target \cite{Gade2007,McDaniel2012,Gade2016R,Gade2016}. 
Here, the extracted one-neutron transfer cross section is higher due to the momentum matching of the well-bound neutron. 
The product of the momentum transfer $q$ and the radius of $^{14}$O nucleus $R$ is around (1--2) $\hbar$ at forward angles, which fits the momentum matching condition \cite{Brink1972,Kay2013}.
Further calculations at 300 MeV/nucleon show
the $qR$ product increases to (3--5) $\hbar$
and the ($\textit{p}$, $\textit{d}$) transfer cross section decreases to about 0.2 mb, negligible compared to the quasifree knockout cross sections~\cite{Atar2018,Aumann2013}.
\begin{figure}[h]
    \centering
    \includegraphics[trim=0cm 0.0cm 0.0cm 0.0cm,clip,width=8.6cm, height=6.0cm]{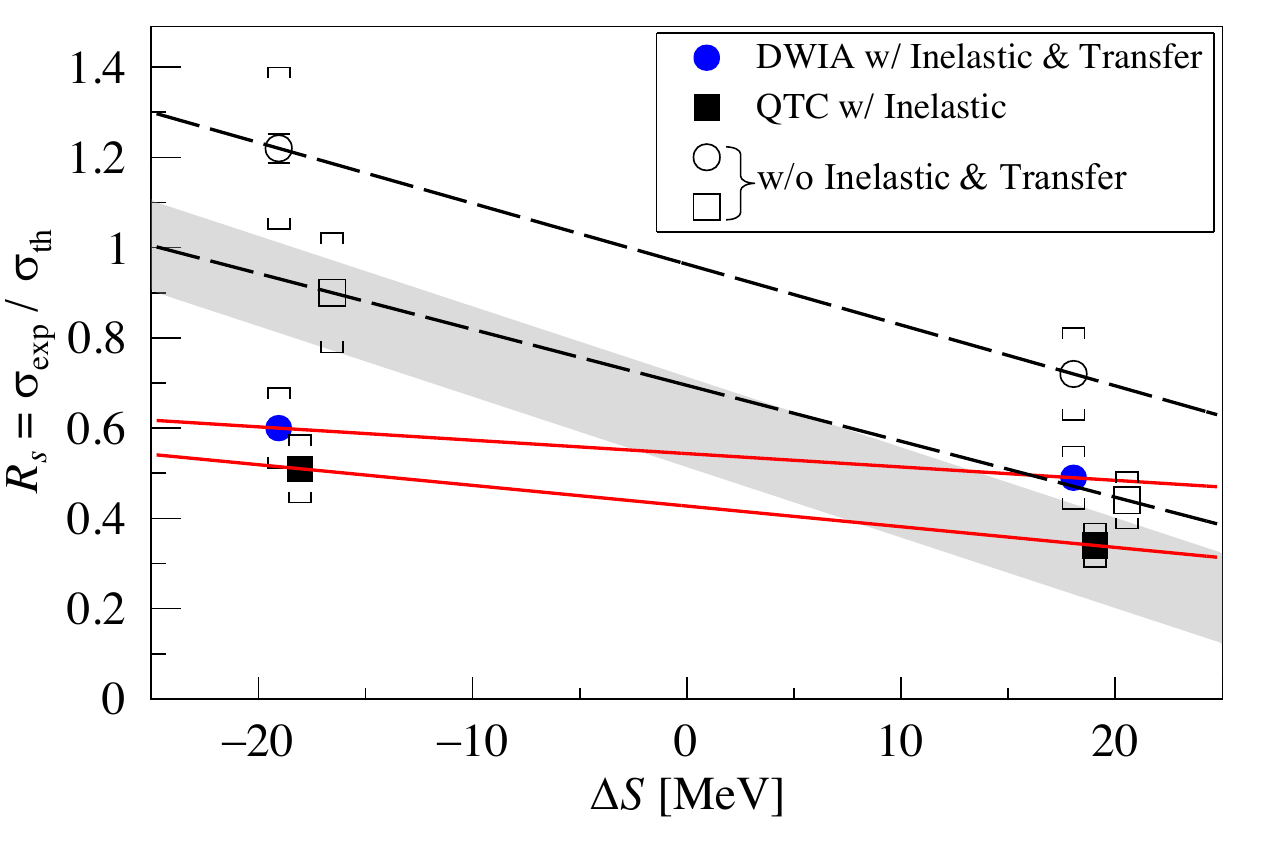}
    \caption{$R_{s}$ as a function of $\Delta S$ from the present work (blue dots and black squares) compared to the trend extracted from Be or C induced nucleon-removal cross sections analyzed with the eikonal model \cite{Gade2008,tostevingade14, tostevingade21} (gray shaded region). 
    The square brackets indicate the total systematic uncertainties.
    Red solid and black dashed lines are shown to guide the eyes.
    }
    \label{fig:Rs} 
\end{figure}
\textcolor{mycolor}{The transfer contribution should thus be assessed for one-nucleon removal reactions
at intermediate energies, especially at energies below 100 MeV/nucleon.
We infer that the one-nucleon removal with a nuclear target may also contain non-negligible transfer contributions, 
where the removed nucleon combines with the target nucleus forming bound or resonance states, depending on the energy and angular momentum matching.}

$R_s$ as a function of $\Delta S$ is shown in Fig.~\ref{fig:Rs}.
Most light-ion-induced nucleon removal $R_s$ lie within a band
with a slope of $-1.6\times10^{-2}$ MeV$^{-1}$ and a half width of 0.1~\cite{Gade2008,tostevingade14,tostevingade21},
as shown by the shaded gray region.
Contrastingly, analyses of low-energy one-nucleon transfer~\cite{jenny2007,flavigny13, Xu2019,Manfredi2021} and 
high-energy quasifree scattering data~\cite{Atar2018,Kawase2018,Gomez2018,Nguyen2019} give slope absolute values of ($10^{-3}$ -- $10^{-5}$) MeV$^{-1}$.
By considering the two datasets of the present work, we obtain a slope of $-3.0(5)(5)\times10^{-3}$ MeV$^{-1}$ 
when the DWIA together with the inelastic and transfer calculations are considered, and of $-4.6(4)(7) \times10^{-3}$ MeV$^{-1}$ 
when the QTC and the inelastic scattering are considered. 
Both slopes are negative and their absolute values are almost zero, indicating $R_{s}$ have a weak $\Delta S$ dependence. 
For comparison, we also extract the $R_s$ if the inelastic scattering and nucleon transfer are neglected in cross-section calculations.
The resulting $R_s$ slopes in the absolute values are 3--5 times larger and look compatible with the strong $\Delta S$ dependence indicated by the light-ion-induced nucleon removal.

In summary, we have reported on the first study of the one-nucleon removal reactions from a large Fermi-surface asymmetric nucleus $^{14}$O ($\Delta S$ = $\pm$18.6 MeV) using a proton target at $\sim$100 MeV/nucleon, a widely used energy regime for rare-isotope studies. 
The measured cross sections and PMDs were compared to the state-of-the-art reaction models, including quasifree knockout, inelastic scattering and nucleon transfer calculations.
In the loosely bound proton removal channel, the ($\textit{p}$, $\textit{p}'$) inelastic scattering and the ($\textit{p}$, 2$\textit{p}$) quasifree knockout are found of almost equal contributions, advocating for an explicit treatment of the inelastic scattering for quantitative interpretation of loosely bound nucleon removal cross sections. 
A highly asymmetric PMD was observed in the deeply bound neutron removal channel, which was reproduced by combining the ($\textit{p}$, $\textit{pn}$) knockout component from DWIA calculations and the ($\emph{p}$, $\emph{d}$) transfer component from DWBA calculations. We observed a distinctive contribution of $\sim$30\% in the high-momentum part of the residue PMD from the deeply bound nucleon stripping ($\textit{p}$, $\textit{d}$) transfer reaction, usually not considered at such beam energies.
The reduction factors extracted from the present two new datasets show a weak $\Delta S$ dependence, 
which is at tension with the eikonal analysis of light-ion-induced knockout reactions.
The extracted dependence 
becomes markedly steeper if the inelastic scattering and nucleon transfer contributions are ignored, 
suggesting that these processes should be considered in analyses of inclusive one-nucleon removal cross sections measured at intermediate energies for quantitative investigation of single-particle strengths and correlations in atomic nuclei.

We are grateful to the RIKEN Nishina Center accelerator staff for providing the stable and high-intensity $^{18}$O beam 
and to the BigRIPS team for the smooth operation of the secondary beam.
This work was supported by the Deutsche Forschungsgemeinschaft (DFG, German Research Foundation)—Projektnummer 279384907–SFB 1245. Y. L. S. and A. O. acknowledge the support from the Alexander von Humboldt foundation. Y. L. S. acknowledges the support of Marie Sk\l{}odowska-Curie Individual Fellowship (H2020-MSCA-IF-2015-705023) from the European Union 
and the support from the Helmholtz International Center for FAIR.
K. O. acknowledges the support by Grant-in-Aid for Scientific Research JP21H00125.
M. G. R.\ acknowledges financial support by MCIN\slash AEI \slash 10.13039 \slash 501100011033 under I+D+i project No.\ PID2020-114687GB-I00, by the Consejer\'{\i}a de Econom\'{\i}a, Conocimiento, Empresas y Universidad, Junta de Andaluc\'{\i}a (Spain) and ``ERDF-A Way of Making Europe'' under PAIDI 2020 project No.\ P20\_01247, and by the European Social Fund and Junta de Andalucía (PAIDI 2020) under grant no. DOC-01006.
C. X. Y. acknowledges Guangdong Major Project of Basic and Applied Basic Research (2021B0301030006).
J. L. acknowledges the support from Research Grants Council (RGC) of Hong Kong (GRF-17303717). 
T. N. acknowledges the JSPS Kakenhi Grants No. JP16H02179, JP18H05404.
Y. T. acknowledges the JSPS Grant-in-Aid for Scientific Research Grants No. JP21H01114.
H. N. L. acknowledges the Fundamental Research Funds for the Central Universities of China.
This work was supported by the Institute for Basic Science (IBS-R031-D1).
B. A. B. was supported by NSF grant PHY-2110365.
Y. Satou is thanked for his help with the inelastic scattering calculations. We thank T. Aumann, C. A. Bertulani and M. Duer for their valuable comments on the manuscript.

\bibliography{References}
\end{document}


\title{Supplemental Material for ``Multiple mechanisms in proton-induced nucleon removal at $\sim$100 MeV/nucleon"}

\newcommand{\akth}{        \affiliation{Department of Physics, Royal Institute of Technology, SE-10691 Stockholm, Sweden}}
\newcommand{\aatomki}{     \affiliation{MTA Atomki, P.O. Box 51, Debrecen H-4001, Hungary}}
\newcommand{\abeijing}{    \affiliation{State Key Laboratory of Nuclear Physics and Technology, Peking University, Beijing 100871, P.R. China}}
\newcommand{\acaen}{       \affiliation{LPC Caen, ENSICAEN, Université de Caen, CNRS/IN2P3, F-14050 Caen, France}}
\newcommand{\acea}{        \affiliation{IRFU, CEA, Universit\'e Paris-Saclay, F-91191 Gif-sur-Yvette, France}}
\newcommand{\acns}{        \affiliation{Center for Nuclear Study, University of Tokyo, RIKEN campus, Wako, Saitama 351-0198, Japan}}
\newcommand{\aewha}{       \affiliation{Department of Physics, Ewha Womans University, Seoul, South Korea}}
\newcommand{\agsi}{        \affiliation{GSI Helmoltzzentrum f\"ur Schwerionenforschung GmbH, Planckstr. 1, 64291 Darmstadt, Germany}}
\newcommand{\ahku}{        \affiliation{Department of Physics, The University of Hong Kong, Pokfulam, Hong Kong}}
\newcommand{\ainst}{       \affiliation{Institute for Nuclear Science \& Technology, VINATOM, 179 Hoang Quoc Viet, Cau Giay, Hanoi, Vietnam}}
\newcommand{\aipno}{       \affiliation{Université Paris-Saclay, CNRS/IN2P3, IJCLab, 91405 Orsay, France}}
\newcommand{\akoeln}{      \affiliation{Institut f\"ur Kernphysik, Universit\"at zu K\"oln, D-50937 Cologne, Germany}}
\newcommand{\alanzhou}{    \affiliation{Institute of Modern Physics, Chinese Academy of Sciences, Lanzhou, China}}
\newcommand{\amadrid}{     \affiliation{Instituto de Estructura de la Materia, CSIC, E-28006 Madrid, Spain}}
\newcommand{\aorsay}{      \affiliation{CSNSM, CNRS/IN2P3, Universit\'e Paris-Sud, F-91405 Orsay Campus, France}}
\newcommand{\aoslo}{       \affiliation{Department of Physics, University of Oslo, N-0316 Oslo, Norway}}
\newcommand{\ariken}{      \affiliation{RIKEN Nishina Center, 2-1 Hirosawa, Wako, Saitama 351-0198, Japan}}
\newcommand{\arikkyo}{     \affiliation{Department of Physics, Rikkyo University, 3-34-1 Nishi-Ikebukuro, Toshima, Tokyo 172-8501, Japan}}
\newcommand{\atitech}{     \affiliation{Department of Physics, Tokyo Institute of Technology, 2-12-1 O-Okayama, Meguro, Tokyo, 152-8551, Japan}}
\newcommand{\atohoku}{     \affiliation{Department of Physics, Tohoku University, Sendai 980-8578, Japan}}
\newcommand{\atud}{        \affiliation{Institut f\"ur Kernphysik, Technische Universit\"at Darmstadt, 64289 Darmstadt, Germany}}
\newcommand{\aunal}{       \affiliation{Universidad Nacional de Colombia, Carr. 30 No. 45-03, Bogot\'a, Colombia}}
\newcommand{\atokyo}{      \affiliation{Department of Physics, University of Tokyo, 7-3-1 Hongo, Bunkyo, Tokyo 113-0033, Japan}}
\newcommand{\aoak}{        \affiliation{Physics Division, Oak Ridge National Laboratory, Oak Ridge, Tennessee 37831, USA}}
\newcommand{\atennessee}{  \affiliation{Department of Physics and Astronomy, University of Tennessee, Knoxville, Tennessee 37996, USA}}
\newcommand{\atriumf}{     \affiliation{TRIUMF 4004 Wesbrook Mall, Vancouver, British Columbia V6T 2A3, Canada}}
\newcommand{\amaxplank}{   \affiliation{Max-Planck-Institut f\"ur Kernphysik, Saupfercheckweg 1, 69117 Heidelberg, Germany}}
\newcommand{\arcnp}{       \affiliation{Research Center for Nuclear Physics (RCNP), Osaka University, Ibaraki 567-0047, Japan}}
\newcommand{\ajaea}{       \affiliation{Advanced Science Research Center, Japan Atomic Energy Agency, Tokai, Ibaraki 319-1195, Japan}}
\newcommand{\asurrey}{     \affiliation{Department of Physics, University of Surrey, Guildford GU2 7XH, UK}}
\newcommand{\aku}{         \affiliation{KU Leuven, Instituut voor Kern- en Stralingsfysica, 3001 Leuven, Belgium}}
\newcommand{\astrasbourg}{ \affiliation{Universit\' de Strasbourg, IPHC, 67037 Strasbourg Cedex, France}}
\newcommand{\atokuyama}{   \affiliation{National Institute of Technology, Tokuyama College, Shunan 745-8585, Japan}}
\newcommand{\apku}{        \affiliation{State Key Laboratory of Nuclear Physics and Technology, School of Physics, Peking University, Beijing 100871, China}}
\newcommand{\amiyazaki}{   \affiliation{Department of Applied Physics, University of Miyazaki, Gakuen-Kibanadai-Nishi 1-1, Miyazaki 889-2192, Japan}}
\newcommand{\akyushu}{     \affiliation{Department of Physics, Kyushu University, Fukuoka 812-8581, Japan}}
\newcommand{\atum}{        \affiliation{Department of Physics, Technische Universit\"at M\"unchen, James-Franck-Stra{\ss}e 1, 85748 Garching, Germany}}
\newcommand{\akyoto}{      \affiliation{Department of Physics, Kyoto University, Kitashirakawa, Sakyo, Kyoto 606-8502, Japan}}
\newcommand{\ausa}{      \affiliation{Departamento de F\'isica At\'omica, Molecular y Nuclear, Facultad de F\'isica, Universidad de Sevilla, Apartado 1065, E-41080 Sevilla, Spain}}
\newcommand{\azsd}{        \affiliation{Sino-French Institute of Nuclear Engineering and Technology, Sun Yat-Sen University, Zhuhai, 519082, Guangdong, China}}
\newcommand{\abnu}{        \affiliation{Key Laboratory of Beam Technology and Material Modification of Ministry of Education,
College of Nuclear Science and Technology, Beijing Normal University, Beijing 100875, China}}
\newcommand{\anscl}{        \affiliation{Department of Physics and Astronomy and the Facility for Rare Isotope Beams, Michigan State University, East Lansing, Michigan 48824-1321, USA }}
\newcommand{\acens}{         \affiliation{Center for Exotic Nuclear Studies, Institute for Basic Science, Daejeon 34126, Republic of Korea}}
\newcommand{\ahoria}{       \affiliation{Horia Hulubei National Institute for R$\&$D in Physics and Nuclear Engineering, IFIN-HH, 077125 Bucure\c{s}ti-M\u{a}gurele, Romania}}
\newcommand{\abuch}{        \affiliation{Doctoral School of Physics, University of Bucharest, 077125 Bucure\c{s}ti-M\u{a}gurele, Romania}}

\author{T.~Pohl} \atud
\author{Y.~L.~Sun} \email{ysun@ikp.tu-darmstadt.de} \atud \acea \ahku
\author{A.~Obertelli} \atud \acea
\author{J.~Lee} \ahku
\author{M. G\'omez-Ramos} \ausa
\author{K.~Ogata} \akyushu \arcnp
\author{K.~Yoshida} \ajaea
\author{B.~S.~Cai}  \azsd
\author{C.~X.~Yuan}  \azsd
\author{B.~A. Brown}  \anscl
\author{H.~Baba} \ariken
\author{D.~Beaumel} \aipno
\author{A.~Corsi}\acea
\author{J.~Gao} \ariken \apku
\author{J.~Gibelin} \acaen
\author{A.~Gillibert} \acea
\author{K.~I.~Hahn} \aewha \acens
\author{T.~Isobe} \ariken
\author{D.~Kim} \aewha \acens
\author{Y.~Kondo}\atitech
\author{T.~Kobayashi} \atohoku
\author{Y.~Kubota} \ariken \acns
\author{P.~Li} \ahku
\author{P.~Liang} \ahku
\author{H.~N.~Liu} \atud \acea \abnu 
\author{J.~Liu} \ahku
\author{T.~Lokotko} \ahku
\author{F.~M.~Marqu\'{e}s} \acaen
\author{Y.~Matsuda} \atohoku
\author{T.~Motobayashi} \ariken
\author{T.~Nakamura} \atitech
\author{N.~A.~Orr} \acaen
\author{H.~Otsu} \ariken
\author{V.~Panin} \ariken \acea
\author{S.~Y.~Park} \aewha \ariken
\author{S.~Sakaguchi} \akyushu
\author{M.~Sasano} \ariken
\author{H.~Sato} \ariken
\author{H.~Sakurai} \ariken \atokyo
\author{Y.~Shimizu} \ariken
\author{A.~I.~Stefanescu} \ahoria \abuch \ariken
\author{L.~Stuhl} \ariken \acens
\author{D.~Suzuki} \ariken
\author{Y.~Togano} \atitech \ariken \arikkyo
\author{D.~Tudor} \ahoria \abuch \ariken
\author{T.~Uesaka} \ariken
\author{H.~Wang} \ariken
\author{X.~Xu} \ahku
\author{Z.~H.~Yang} \ariken
\author{K.~Yoneda} \ariken
\author{J.~Zenihiro} \ariken

\date{\today}
\maketitle

\section{DWIA and QTC calculations}
In the DWIA calculations, we employed the folding potential with the Melbourne G-matrix interaction to calculate the distorted waves. As for the transition process, we adopted the nucleon-nucleon cross section calculated with the Franey-Love interaction~\cite{Franey1985}. Perey correction~\cite{Perey1962} for the non-locality was applied both for the bound-state and scattering wave functions. 
Energy dependence of the optical potentials was taken into account by using the scattering energy of the emitted nucleons. 
In the QTC calculations, the $\textit{p} + N$ + $^{13}$O/$^{13}$N three-body final state was expressed in a basis of discretized continuum states of the $\textit{p}$ + $\textit{N}$ system. Microscopic JLM optical potential~\cite{jlm77} was employed for the distortion of the incident and outgoing channels. Both the G-matrix folding potential and JLM potential were found to reproduce well the  experimental differential cross section for $\textit{p}$ + $^{16}$O elastic scattering at 65 MeV. The single-particle wave function of the knocked-out nucleon in both DWIA and QTC was obtained by solving the Schr\"odinger equation using a Woods-Saxon potential with an adjusted depth to result in the correct binding energy of the knocked-out nucleon.
The core radius and the diffuseness of the Woods-Saxon potential were fixed at 3.27 fm and 0.7 fm, respectively.

In the QTC calculation for the $(p, d)$ transfer,
the $\textit{d}$--$^{13}$O potential was calculated with the Johnson-Soper prescription \cite{Johnson1970}, in which the $\textit{p}$--$^{13}$O and $\textit{n}$--$^{13}$O folding potential at half of the kinetic energy of the deuteron was adopted. 
\textcolor{black}{The deuteron wave function was obtained using the Reid93 interaction~\cite{Stoks1994},
which reproduces its experimental $\textit{s}$- and $\textit{d}$-wave components.}
The obtained single-particle cross section ($\sigma\rm_{sp}$) for the transfer reaction was 3 mb. The uncertainty was about 1 mb, estimated by using the JLM $\textit{d}$--$^{13}$O potential and by varying the interaction used for the deuteron wave function.

\begin{figure}[h]
    \centering
    \includegraphics[trim=0cm 0.2cm 0.8cm 1.2cm,clip,width=9.0cm,height=8.0cm]{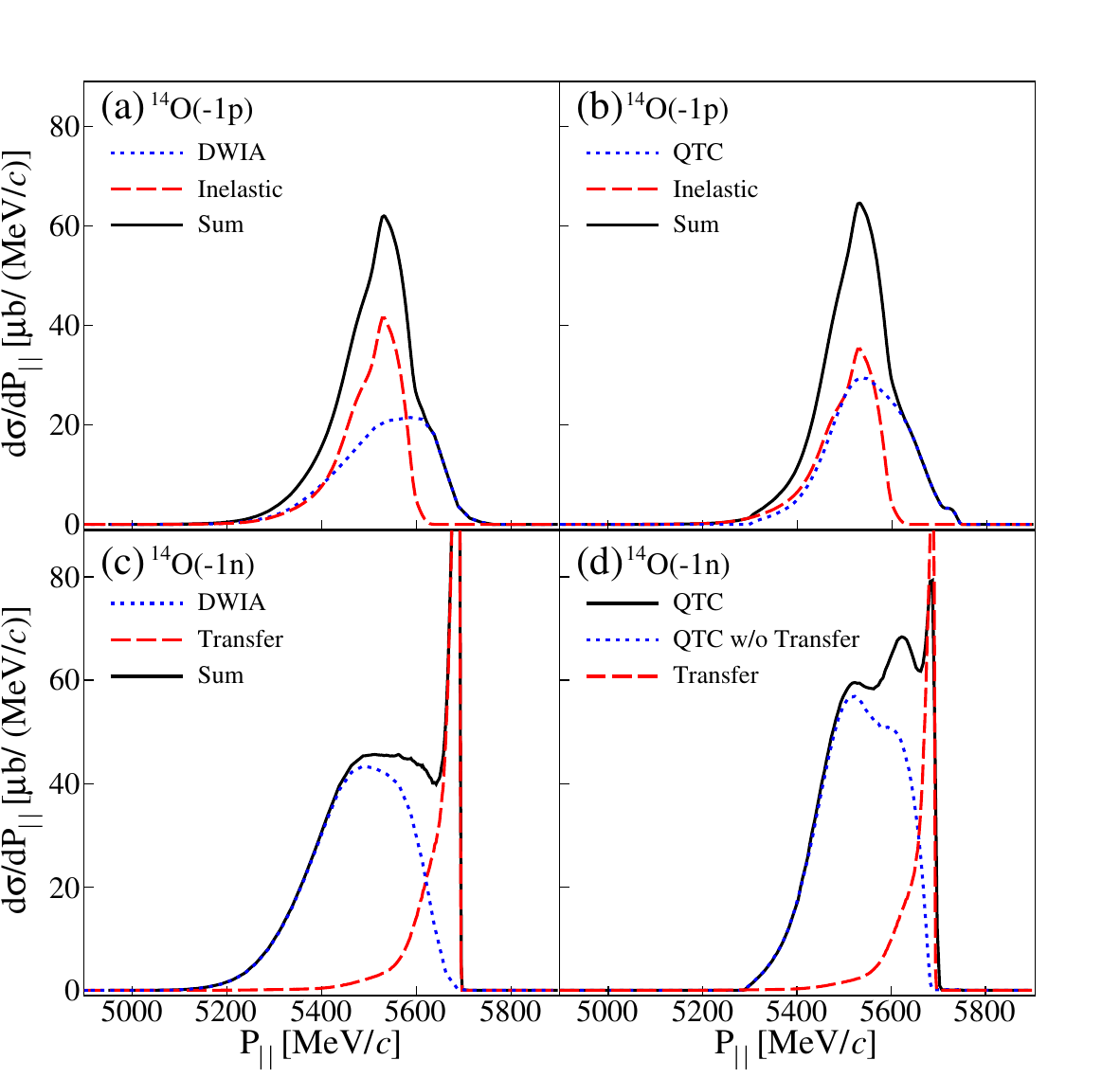}
    \caption{Theoretical parallel momentum distributions of $^{13}$N and $^{13}$O from the one-nucleon removal of $^{14}$O at 94 MeV/nucleon.
    Panels (a)-(c) show DWIA and QTC calculations (blue-dotted lines), 
    with additional contributions from inelastic excitation for $^{13}$N and ($\textit{p}$, $\textit{d}$) transfer for $^{13}$O (red-dashed lines). 
    Panel (d) displays the QTC calculation as a black-solid line, while the calculation without (w/o) ($\textit{p}$, $\textit{d}$) transfer is represented by a blue-dotted line.
    Theoretical distributions have not been convoluted with the experimental response, but the integrals have been normalized to the experimental cross sections. Note that the experimental response will dampen and smooth out the sharp ($\textit{p}$, $\textit{d}$) transfer peak.
    }
    \label{fig:PMD} 
\end{figure}

The calculated parallel momentum spectra are shown in Fig.~\ref{fig:PMD}.

\section{Inelastic excitation calculations}
The inelastic excitation cross section was calculated using the microscopic DWIA reaction model~\cite{DW81}, that has been validated for the inelastic scattering of proton on the $^{12}$C target. The structure input for the calculation is the one-body transition densities (OBTDs) from the shell model calculations. We adopted the Franey-Love nucleon-nucleon effective interaction to calculate the transition amplitudes. The Koning-Delaroche phenomenological optical potential~\cite{Koning03} was used to generate the distorted waves. The calculation reproduce reasonably well the differential cross section of the $2_{1}^{+}$ excitation of $^{12}$C measured at 120 MeV~\cite{Comfort1981}. We have considered eight excited states of $^{14}$O ($0_{2}^{+},0_{3}^{+},2_{1}^{+}, 2_{2}^{+},0_{1}^{-},1_{1}^{-},2_{1}^{-}, 3_{1}^{-}$) that could decay via one proton emission to $^{13}$N~\cite{Charity2019}, resulting in a total inelastic cross section of 9 mb. The dominant contribution originated from the $2_{1}^{+}, 1_{1}^{-}$ and $3_{1}^{-}$ excitations. 
The uncertainty is about 1mb, estimated by performing calculations using 
different effective interactions
for the OBTDs, as well as by using the M3Y nucleon-nucleon interaction, that takes the medium effects into account~\cite{Bertsch1977}.\\

\section{Calculations of spectroscopic factors and $\sigma\rm_{th}$}
\textcolor{black}{Spectroscopic factors (SFs) used in this work were obtained from shell-model calculations 
performed in the $\textit{psd}$-model space with the YSOX interaction~\cite{Yuan12} limited to 5 $\hbar\omega$ excitation using the KSHELL code~\cite{Shimizu19}. The single-particle energy of the $\pi$1\textit{s}$_{1/2}$ orbit was decreased by 0.375 MeV to have a good reproduction of the low-lying energy levels of $^{14}$O. The resulting SFs are 1.58 and 3.42 for $^{13}$N and $^{13}$O, respectively.
OXBASH shell-model calculations \cite{oxbash2004} with the commonly used WBT and WBP interactions \cite{WBT1992}
predict slightly larger SFs of 1.82 and 3.72.}

 The theoretical cross section $\sigma\rm_{th}$ for the loosely-bound proton removal is the sum of the ($\textit{p}$, 2$\textit{p}$) knockout cross section ($\sigma\rm_{sp}$ $\times$ SF $\times$ $A/(A-1)$) and the inelastic excitation cross section. Note that $A/(A-1)$ is the center-of-mass correction factor~\cite{Dieperink1974} and $A$ is the mass number of the projectile, $\textit{i.e.}$, 14 in this case.

 Since the ($\textit{p}$, $\textit{d}$) transfer process is considered in the QTC formalism but not in the DWIA, we calculated the $\sigma\rm_{th}$ for the neutron removal as: 
1) the $\sigma\rm_{sp}$ from the DWIA and the ($\textit{p}$, $\textit{d}$) transfer multiplied by the SF $\times$ $A/(A-1)$ factor.
2) the $\sigma\rm_{sp}$ from the QTC calculation multiplied by the SF $\times$ $A/(A-1)$ factor.

\bibliography{References}